\documentclass[%
 reprint,
 superscriptaddress,
 amsmath,amssymb,
 aps,
]{revtex4-2}

\usepackage{graphicx}% Include figure files
\usepackage{dcolumn}% Align table columns on decimal point
\usepackage{bm}% bold math
\usepackage{siunitx}
\usepackage{xcolor}
\usepackage{soul}
\setstcolor{blue}

\begin{document}

\preprint{APS/123-QED}

\title{Meta-cavity Quantum Electrodynamics}

\author{Xueshi Li}
\thanks{These authors contribute equally to this research.}
\affiliation{Institute for Quantum Science and Technology, National University of Defense Technology, Changsha, 410073, China}

\author{Ziwei Wang}
\thanks{These authors contribute equally to this research.}
\affiliation{State Key Laboratory of Photonics and Communications, School of Physics and Astronomy, Shanghai Jiao Tong University, Shanghai, 200240, China}

\author{Yan Chen}
\email{chenyan@nudt.edu.cn}
\thanks{These authors contribute equally to this research.}
\affiliation{Institute for Quantum Science and Technology, National University of Defense Technology, Changsha, 410073, China}

\author{Dong Liu}
\affiliation{State Key Laboratory of Optoelectronic Materials and Technologies, Sun Yat-sen University, Guangzhou, 510275, China}

\author{Kaili Xiong}
\affiliation{Institute for Quantum Science and Technology, National University of Defense Technology, Changsha, 410073, China}

\author{Guangfeng Wang}
\affiliation{State Key Laboratory of Photonics and Communications, School of Physics and Astronomy, Shanghai Jiao Tong University, Shanghai, 200240, China}

\author{Jiantao Ma}
\affiliation{State Key Laboratory of Optoelectronic Materials and Technologies, Sun Yat-sen University, Guangzhou, 510275, China}

\author{Ying Yu}
\affiliation{State Key Laboratory of Optoelectronic Materials and Technologies, Sun Yat-sen University, Guangzhou, 510275, China}

\author{Jiawei Wang}
\affiliation{School of Integrated Circuits, Harbin Institute of Technology (Shenzhen), Shenzhen 518055, China}

\author{Zhanling Wang}
\affiliation{College of Electronic Science and Technology, National University of Defense Technology, Changsha, 410073, China}

\author{Xiao Li}
\affiliation{College of Advanced Interdisciplinary Studies, National University of Defense Technology, Changsha, 410073, China}

\author{Xianfeng Chen}
\affiliation{State Key Laboratory of Photonics and Communications, School of Physics and Astronomy, Shanghai Jiao Tong University, Shanghai, 200240, China}
\affiliation{Shanghai Research Center for Quantum Sciences, Shanghai, 201315, China}
\affiliation{Collaborative Innovation Center of Light Manipulations and Applications, Shandong Normal University, Jinan, 250358, China}

\author{Erez Hasman}
\affiliation{Atomic-Scale Photonics Laboratory, Russell Berrie Nanotechnology Institute, and Helen Diller Quantum Center, Technion Israel Institute of Technology, Haifa, 3200003, Israel}

\author{Bo Wang}
\email{wangbo89@sjtu.edu.cn}
\affiliation{State Key Laboratory of Photonics and Communications, School of Physics and Astronomy, Shanghai Jiao Tong University, Shanghai, 200240, China}

\author{Jin Liu}
\email{liujin23@mail.sysu.edu.cn}
\affiliation{State Key Laboratory of Optoelectronic Materials and Technologies, Sun Yat-sen University, Guangzhou, 510275, China}

\author{Tian Jiang}
\email{tjiang@nudt.edu.cn}
\affiliation{Institute for Quantum Science and Technology, National University of Defense Technology, Changsha, 410073, China}
\affiliation{Hunan Research Center of the Basic Discipline for Physical States, Changsha, 410073, China}

\date{\today}

\begin{abstract}
Cavity quantum electrodynamics (cQED) harnesses light-matter interactions to produce nonclassical light states. However, a fundamental challenge lies in simultaneously achieving Purcell enhancement and tailored wavefront control within a single cavity, due to conflicting resonator requirements. Here, we overcome this limitation by demonstrating triggered single-photon emission with customizable wavefronts from semiconductor quantum dots embedded in geometric-phase meta-cavities. These monolithic devices - only \SI{200}{nm} thick - deliver Purcell-enhanced emission alongside spin-momentum-locked radiation, vortex beams, and holographic patterns. The meta-atom lattice provides high-Q optical confinement, while spatially modulated orientations enable efficient outcoupling of photons with designed states. This work establishes a new paradigm for intrinsically multiplexing metasurface-based wavefront shaping with cQED, enabling high-performance quantum light sources from subwavelength-scale monolithic platforms.
\end{abstract}

\maketitle

\begin{figure*}
\centering
\includegraphics[width=0.8\textwidth]{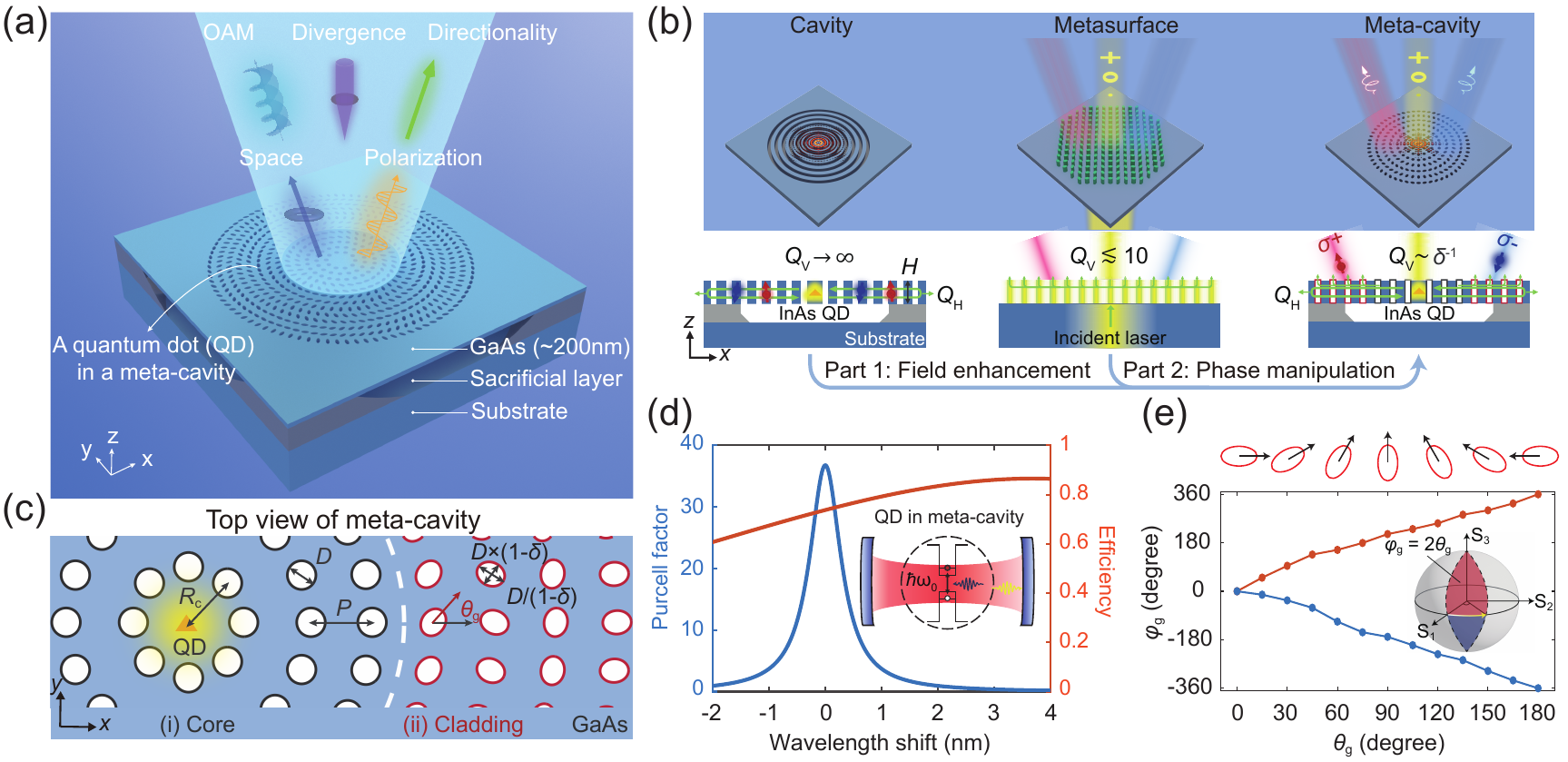}
\caption{\textbf{Working principles of the quantum electrodynamics in a GP meta-cavity.} (\textbf{a}) Schematic of the monolithic meta-cavity. (\textbf{b}) Roadmap of the meta-cavity. Left panel: A typical first-order circular Bragg cavity. Middle panel: A typical GP meta-surface. Right panel: A meta-cavity. (\textbf{c}) Top-view of the meta-cavity. QD is placed in the center of the cavity. (\textbf{d}) A typical simulated Purcell factor and radiation efficiency for a meta-cavity. (\textbf{e}) GP $\varphi_g$ obtained from orientations $\theta_g$ of the elliptical meta-atoms in the cladding region.}
\label{fig:1}
\end{figure*}

\section{Introduction}
Quantum light sources are indispensable for a wide range of advanced technologies, such as quantum key distribution communication \cite{scarani2009}, quantum simulation \cite{aspuru2012}, imaging \cite{defienne2024}, and sensing \cite{pirandola2018}. A key requirement in developing practical quantum light sources is to achieve deterministic single photons with well-defined photonic states \cite{somaschi2016,he2017,tomm2021}. However, the spontaneous emission of a single quantum emitter usually gives rise to randomly orientated dipole radiation, which is statistically unpolarized and omnidirectional. The control of single photon's properties therefore necessitates well-designed emitter-cavity heterostructures, wherein the quantum emitter should be precisely embedded in a cavity, such as a photonic crystal \cite{barik2018}, micropillars \cite{somaschi2016,tomm2021,wang2019,santori2002,ding2016,chen2025} or distributed Bragg gratings \cite{wang2019,chen2025,liu2019,wang2019b}. By combining resonant excitation with Purcell-enhanced cavities, the generated single photons can be efficiently extracted into the far field with deterministic photon properties. Achieving strong Purcell enhancement requires cavities with both high-quality factors (Q) and wavelength-scale mode volumes \cite{painter1999,yoshie2001,akahane2003,yoshie2004,lalanne2008,liu2018,yan2025,Chen2025b}. However, these conditions fundamentally conflict with wavefront shaping, as desired spatial light manipulation necessitates meta-structures spanning multiple wavelengths \cite{pors2013,yu2014,zheng2015,wang2016,maguid2016,zeng2025}. Therefore, the customizing of photon properties is usually achieved by using an additional optical device, such as a meta-surface, that is placed in the vicinity or even far away from the emitted photons \cite{stav2018,huang2019,bao2020,li2020,xie2020,solntsev2021,komisar2023}. The meta-surfaces are constructed of arrays of customized nano-antennas that can be used to control the amplitude, phase and polarization states of light at the subwavelength resolution \cite{bomzon2001,yu2011,qu2015,lin2014,sun2012,ha2018}. Particularly, geometric phase (GP) meta-surfaces are constructed of anisotropic nano-antennas with space-variant orientations, which allow the polarization evolution of light on the Poincar\'{e} sphere to generate spin-dependent phases \cite{bomzon2001,pancharatnam1956,berry1984}. Previously, GP meta-surfaces have been used to achieve photonic Rashba effects \cite{rong2020,rong2023,duan2023}, spin-orbital entanglements \cite{stav2018}, high-order orbital states \cite{liu2023}, and directional single-photon emission and polarization control \cite{bao2020}. Because of the low Q-factor, these meta-surfaces offer customized wavefront control of photons without Purcell enhancement.

Recently, the realization of high Q-factor meta-surfaces has been investigated for nonlocal meta-surfaces based on guided-mode resonators \cite{song2021} and photonic crystals \cite{overvig2020,huang2023,nolen2024,qin2025}. The nonlocal meta-surfaces simultaneously serve as a meta-surface and a photonic crystal to support high Q-factor resonant modes with customized spin-dependent radiation channels. In particular, nonlocal meta-surfaces with a GP mechanism have been demonstrated to boost the emission efficiency of different on-chip dipole emitters, such as colloidal quantum dots (QDs) \cite{rong2020} and two-dimensional transition-metal dichalcogenides \cite{rong2023,duan2023}. However, these GP devices are based on the bulky photonic crystal modes to stimulate many emitters on a large area. Therefore, they are not suitable for quantum light emission.

In this work, we demonstrate triggered single photon emission from a InAs-QDs-integrated GP meta-cavity. Our design integrates a single QD in the center of a circular GP meta-cavity that incorporates space-variant GP meta-atoms. We employ a defect mode of circular photonic crystal, enhancing the localization of the optical modes for single quantum emitters, and optimizing the Purcell factor ($F_p$) with reduced mode volume ($V_{\text{mode}}$). By tailoring the geometric parameters of the meta-atoms, such as their size and anisotropic properties, we achieve precise control over wavefront shaping while optimizing photon collection efficiency. This unique approach enables simultaneous Purcell enhancement (experimentally $F_p \sim 9.7$) with multi-dimensional control over spin-orbit interactions, orbital angular momentum (OAM), and holographic wavefront shaping of emitted photons in a single, ultrathin structure with an effective thickness of only \SI{200}{nm}, a paradigm we term "Meta-cavity Quantum Electrodynamics (Meta-c QED)". This monolithic design eliminates the need for cascading cavities and meta-surfaces, addressing the challenges of complexity and scalability inherent in previous approaches.

\section{Results}
\subsection{Working principle of monolithic meta-cavities}
Figure \ref{fig:1}(a) presents the artistic sketch of the monolithic meta-cavity. A circular Bragg grating (CBG) cavity can confine light in a very small volume from the first order Bragg condition, it hosts a high-Q resonant mode ($Q_v \rightarrow \infty$) and allows only horizontal leakage [Fig. \ref{fig:1}(b), left panel]. Therefore, photons are trapped in the cavity without radiation. On the contrary, a GP meta-surface is highly radiative, and it allows for arbitrary wavefront shaping of an external laser but with a relatively small quality factor ($Q_v \lesssim 10$) [Fig. \ref{fig:1}(b), middle panel]. Our meta-cavity takes advantage of both the defect cavity (local field enhancement) and the meta-surface (phase manipulation). Therefore, the radiation efficiency is enhanced and photons leak into space with customized spatial and polarization properties [Fig. \ref{fig:1}(b), right panel]. The quality factor of the GP meta-cavity can be tuned by $Q_v \sim \delta^{-1}$. Here, $\delta$ is the anisotropy of the perturbed elliptical holes to generate GP $\varphi_g$. Specifically, $D$ is the diameter of the unperturbed circular holes, and $D/(1-\delta)$ and $D\times(1-\delta)$ represent the major and minor axes of the elliptical meta-atoms, respectively [Fig. \ref{fig:1}(c)].

The details of the meta-cavity are introduced in Supplementary Note 1. The radial periodicity of the concentric circles is $P$, with $NP$ being the distance of the $N^{\text{th}}$ concentric circle to the center of the cavity, and $N$ being an integer. For the $N^{\text{th}}$ concentric circle, there are $8\times N$ meta-atoms [Fig. \ref{fig:1}(c)]. This arrangement ensures the GP cavity with a high rotational symmetry  ($C_8$) and a reasonable spatial density of meta-atoms. The meta-atoms are circular or elliptical air holes etched through a thin GaAs layer. For $N \leq N_c$, the meta-atoms are circular holes with a unique diameter of $D$, representing a core region of the cavity [Fig. \ref{fig:1}(c)]. A cladding region is defined for $N > N_c$, wherein the meta-atoms are elliptical holes with a spatially-variant orientation $\theta_g(x, y)$ and an anisotropy parameter $\delta \in [0,1)$. Notably, $\delta$ is the key parameter that is used to balance the trade-off between the resonance and radiation properties of our meta-cavity. Moreover, $R_c$ is the distance from the first circular hole to the center of the cavity, which can significantly affect the Purcell factor, while the diameter $D$ of meta-atoms mainly modulates the resonant wavelength. Detailed analysis regarding the influence of key geometric parameters on the meta-cavity performance and the corresponding robustness against variations is provided in Supplementary Note 2.1 to 2.3.

Theoretically, the meta-cavity supports a resonant mode with a quality factor of several thousands. When spectrally overlapped with a QD, this mode yields Purcell-enhanced emission with a calculated factor of above 35 [Fig. \ref{fig:1}(d)]. The outcoupling efficiency, defined as photons collected in free space, exceeds 70\% across a broadband range (Supplementary Note 2.4). Crucially, the spatially extended resonant mode acquires a GP $\varphi_g = 2\sigma_{\pm}\theta_g$ [Fig. \ref{fig:1}(e)], which is independent of dynamic phase and leaves resonance conditions unaffected (Supplementary Note 2.5). Here, $\sigma_{\pm}$ stands for the spin of photons. Consequently, single-photon wavefronts and far-field radiation patterns are solely governed by the meta-atom-assigned GP $\varphi_g$, enabling multi-dimensional engineering of quantum emission profiles.

\begin{figure}
\centering
\includegraphics[width=\columnwidth]{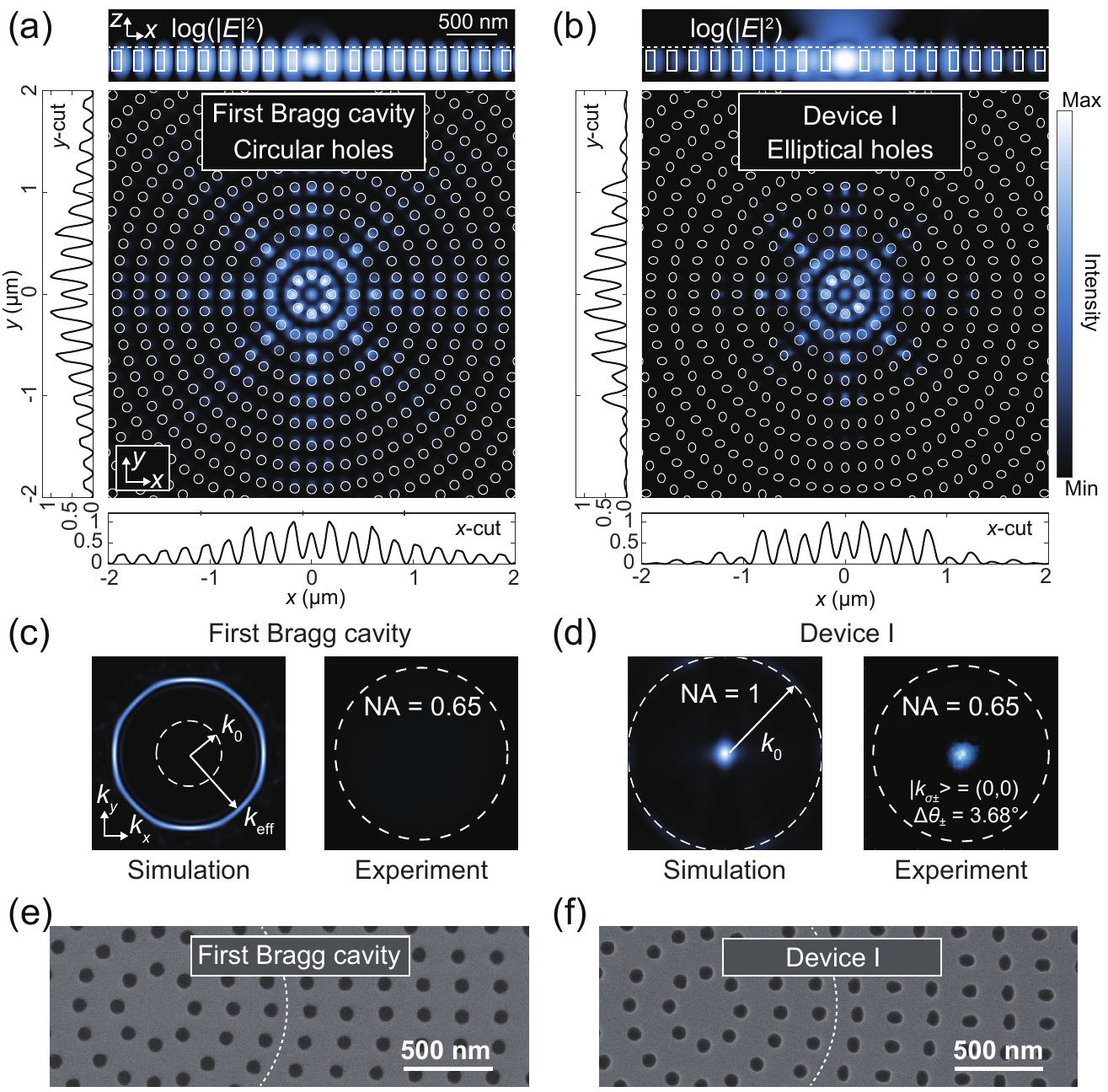}
\caption{\textbf{The properties in the meta-cavities.} (\textbf{a, b}) Simulated near-field intensity distribution of the unperturbed mode (\textbf{a}, $\delta = 0$, $P = \SI{208}{nm}$, $D = \SI{92}{nm}$, $R_c = 0.48\times2P$) and perturbed mode (\textbf{b}, $\delta = 0.2$, $P = \SI{208}{nm}$, $D = \SI{92}{nm}$, $R_c = 0.48\times2P$). Top panel: White dash line represents the near-field position. (\textbf{c, d}) K-space distribution from simulation and experiment for the unperturbed (\textbf{c}) and perturbed (\textbf{d}) modes, where $k_0$ represents NA = 1. The experimental NA is 0.65. (\textbf{e, f}) The SEM images of the meta-cavities, where the meta-atoms are circular (\textbf{e}) and elliptical air holes (\textbf{f}), respectively.}
\label{fig:2}
\end{figure}

\subsection{Device design and fabrication}
Figure \ref{fig:2} presents the simulated and experimental results of our meta-cavities based on a \SI{200}{nm}-thick GaAs film. The resonant modes are stimulated by a broadband circular-polarized dipole emitter located in the center of the cavities. We begin with the case of $\delta = 0$, with all meta-atoms being circular holes for both the core and cladding region [Fig. \ref{fig:2}(a)]. Alongside the strongly confined optical field in the core region, a small portion of the field is also tunneling through the cladding region as evanescent waves fulfilling a first-order Bragg condition, $Pn_{\text{eff}} = \lambda/2$ with $\lambda$ being the working wavelength. This means that the in-plane wave vector of the resonant mode ($\lambda$) is $k_{\text{in}} > k_0 = 2\pi/\lambda$, and $k_{\text{in}}P = \pi$. Here $n_{\text{eff}}$ is an effective refractive index used to describe the in-plane propagation of the wave $\lambda$ in the cavity. 

Furthermore, we slightly deform the cladding meta-atoms into elliptical holes by defining $\delta = 0.2$. Meanwhile, the spatial-variant orientation $\theta_g(x, y)$ of these meta-atoms are designed to create novel radiation channels for photons in the free space with $|k_{\sigma \pm}\rangle = \sigma_{\pm}n_{\text{eff}}k_0\nabla\mathbf{r} - 2\sigma_{\pm}\nabla\theta(\mathbf{r})$, and $\mathbf{r} = (x, y)$. This results in GP meta-cavities, and the radiative photons can be described by $|\Psi\rangle = (|\sigma_{+}\rangle|k_{\sigma +}\rangle \pm |\sigma_{-}\rangle|k_{\sigma -}\rangle)/\sqrt{2}$. Here, $|\sigma_{\pm}\rangle$ denotes the spin-up and spin-down polarization states of photon, and $|k_{\sigma \pm}\rangle$ are the associated momenta in the free space. In Fig. \ref{fig:2}(b), we present the in-plane orientations of the elliptical meta-atoms for $|k_{\sigma \pm}\rangle = (0, 0)$. The spatially varying orientation $\theta_g(x,y)$ of these meta-atoms alternates between 0 and $\pi/2$. Because the GP is a small perturbation from the circular cavity, it only slightly degrades the Q-factor by introducing the radiation channel ($Q_V$), without significantly affecting the nearfield mode. This allows us to predict a high Purcell factor according to $F_p \sim (\lambda/n)^3 Q/V_{\text{mode}}$ (Supplementary Note 2.2).

To demonstrate single-photon emission mediated by the GP cavity, we fabricated freestanding devices comprising a \SI{200}{nm}-thick GaAs membrane with embedded InAs QDs. The meta-cavities stand upon a 600-nm-thick sacrificial layer and a 40-period AlAs/GaAs distributed Bragg reflector, which is used to enhance the collection intensity. Detailed fabrication procedures are provided in Supplementary Note 3. As demonstrated in Fig. \ref{fig:2}(c), the simulated k-space distribution of the resonant mode is a circular pattern with $k_{\text{in}} \approx k_{\text{eff}} \approx 2.24k_0$, meaning that $n_{\text{eff}} \approx 2.24$ and the field is evanescent. The corresponding experimental k-space distribution confirms that photons are trapped in the cavity without radiation due to the high-Q resonant mode under first-order Bragg condition. In contrast, photons are leaked out in the vertical direction because of the GP perturbations caused by the elliptical holes. A Gaussian-like radiation pattern emerges at $|k_{\sigma \pm}\rangle = (0, 0)$ [Fig. \ref{fig:2}(d)], yielding a narrow momentum-space bandwidth of $\Delta k = 0.13$ ($\Delta\theta_{\pm} = 3.68^\circ$). This highly directional emission is ideal for efficient fiber collection ($>$98\% mode overlap with ideal Gaussian profile, Supplementary Note 4). Figure \ref{fig:2}(e) and \ref{fig:2}(f) present the scanning electron microscopy (SEM) images of two typical devices, where the meta-atoms are circular and elliptical air holes, respectively. The diameter of a single GP cavity is about \SI{10}{\micro\meter}.

\subsection{Bright and indistinguishable single-photon emission from meta-cavities}
We firstly examine the characteristics of the GP mediated single-photon source [Device I, Fig. \ref{fig:2}(d)]. The optical setup is described in Supplementary Note 5. As shown in Fig. \ref{fig:3}(a), the observed emission spectrum of the QD exhibits a significant spectral overlap with the cavity resonance mode, with a very narrow spectral width of $\sim\SI{0.02}{nm}$ limited by the resolution of our spectrometer. Under a cross-polarization configuration, the QD was resonantly excited by a pulsed laser with a repetition rate of 80 MHz. The measured optical intensity showed coherent Rabi oscillations, demonstrating coherent population control with a preparation fidelity of 34\% [Inset of Fig. \ref{fig:3}(a)]. To examine the cavity-enhanced emission dynamics, we performed time-resolved lifetime measurements [Fig. \ref{fig:3}(b)]. The plain QD exhibited a radiative lifetime of $\tau = \SI{974.3}{ps}$, while the cavity-coupled QD showed a significantly reduced lifetime of $\tau_c = \SI{102.2}{ps}$. This substantial lifetime reduction corresponds to a Purcell enhancement factor of $F_p = 9.7$, confirming effective cavity quantum electrodynamics coupling.

Single-photon purity was measured using a Hanbury Brown-Twiss (HBT) interferometer [Fig. \ref{fig:3}(c)]. The suppression of coincidence counts at zero-time delay $g^{(2)}(0) = 0.017(2)$ reflects strong antibunching characteristics. Moreover, photon indistinguishability was assessed via a Hong-Ou-Mandel (HOM) interferometer [Fig. \ref{fig:3}(d)]. The raw quantum interference visibility reached $V_{\text{raw}} = 0.730(2)$. After correcting for the residual multiphoton probability, imperfect classical interference, and beam splitter imbalance, a corrected indistinguishability of $V_{\text{cor}} = 0.865(2)$ is obtained \cite{santori2002} (Supplementary Note 6). At $\pi$ pulse, \SI{176}{kHz} photon rate is recorded, corresponding to an extraction efficiency of 31.1\%. Details about the calibration are described in Supplementary Note 4. Notably, both coherent control of the state population with high preparation fidelity and suppression of blinking (i.e., ensuring QD emits exactly one photon per trigger) can be achieved by applying a bias electric field which is compatible with current device design.  \cite{zhai2020,strobel2023}.

\begin{figure}
\centering
\includegraphics[width=\columnwidth]{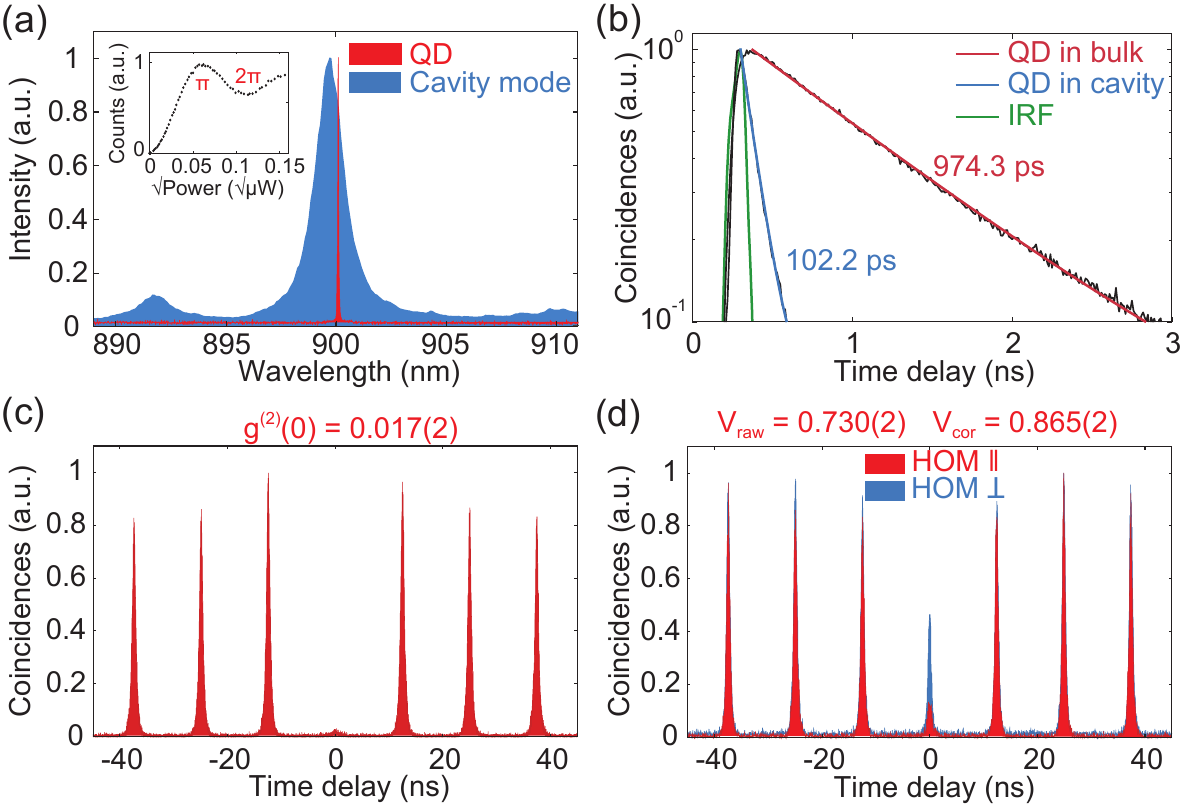}
\caption{\textbf{Experimental characterizations of the meta-cavity single-photon source.} (\textbf{a}) Photoluminescent spectra showing the QD emission (red) spectrally detuned by \SI{0.4}{nm} from the meta-cavity resonance mode (blue). Inset: Coherent Rabi oscillations. (\textbf{b}) Radiative lifetime of the QD. The green curve indicates the instrument response function (IRF). (\textbf{c}) HBT coincidence histogram of single photons. (\textbf{d}) HOM interference histogram of single photons.}
\label{fig:3}
\end{figure}

\subsection{Quantum radiation with multi-dimensionally engineered wavefront control}
\begin{figure}
\centering
\includegraphics[width=\columnwidth]{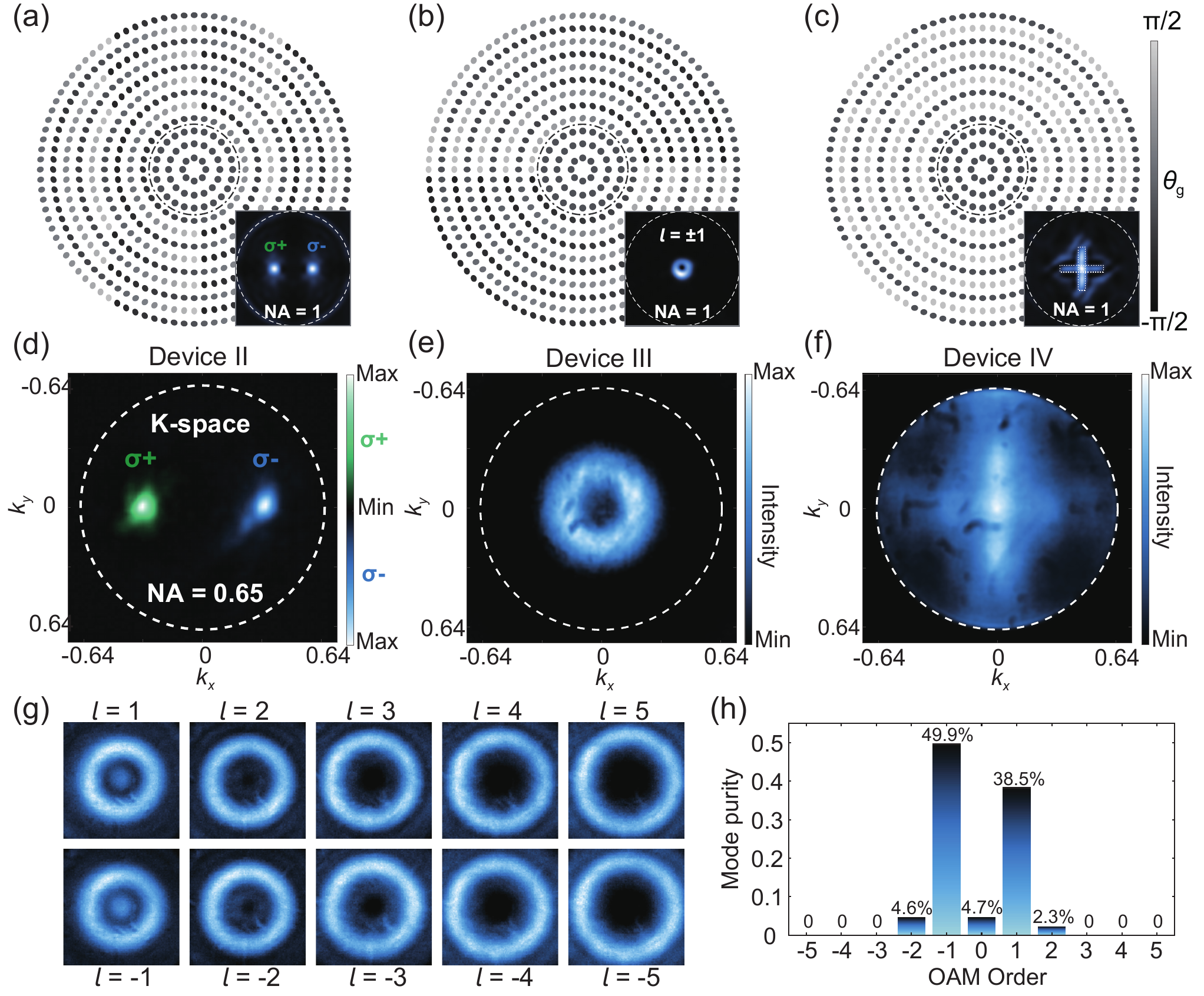}
\caption{\textbf{Experimental demonstrations of photon's spatial state manipulation in GP meta-cavities.} (\textbf{a-c}) The color-code in-plane orientations $\theta_g(x,y)$ of elliptical meta-atoms. The insets are the k-space radiation patterns with light cone marked by white circle. (\textbf{d}) The spin-momentum-locked radiation $|k_{\sigma +}\rangle = (-0.31, 0)$, $|k_{\sigma -}\rangle = (+0.32, 0)$. (\textbf{e}) The OAM emission pattern. (\textbf{f}) The reconstructed hologram of a "+" pattern. (\textbf{g}) The far-field patterns and (\textbf{h}) histogram after phase projection to a series of vortex wave plates with different topological charges of $\ell$.}
\label{fig:4}
\end{figure}

After characterizing the single photons emitted from the meta-cavities, we now introduce more complex functionalities to show the multi-dimensional wavefront control. Without loss of generality, we design representative meta-cavities with distinct functionalities and increasing complexity, as schematically presented in Fig. \ref{fig:4}. Devices with the spin-momentum locking effect, OAM and hologram are demonstrated, respectively. The color of each meta-atom represents the specific GP phases and the insets depict the far field patterns. The detailed design procedure is elaborated in Supplementary Note 1.4. Fig. \ref{fig:4}(a) shows the simulated results of the meta-cavity featuring spin-dependent radiation at the designed $|k_{\sigma \pm}\rangle = (\mp 0.32, 0)$. In Fig. \ref{fig:4}(b), an OAM is generated at $|k_{\sigma\pm}\rangle = (0, 0)$ with topological charge of $\ell = \pm 1$. In Fig. \ref{fig:4}(c), we designed a holographic image of light with a plus sign "+" in the momentum space. Beyond generating showcased photon wavefronts, our meta-cavity platform supports diverse functionalities such as producing optical skyrmions and high-order OAM states (Supplementary Note 2.6).

Experimentally, the corresponding far-field profiles are plotted in the bottom panes of Fig. \ref{fig:4}(d) - (f). Figure \ref{fig:4}(d) demonstrates spin-orbit manipulation of emitted photons. Momentum-space radiation characterization reveals two Gaussian-like far-field spots at diffraction angles of $-2.58^\circ$ and $+2.42^\circ$, respectively. $|\sigma_{+}\rangle$ and $|\sigma_{-}\rangle$ circular polarization states correlate exclusively with $|k_{\sigma+}\rangle$ and $|k_{\sigma-}\rangle$ momentum states. Fig. \ref{fig:4}(e) displays the measured OAM emission pattern. The far-field patterns after phase projection are plotted in Fig. \ref{fig:4}(g), which reveals solid-center lobes exclusively for the first-order mode, while higher-order modes exhibit donut-shaped profiles, confirming successful generation of the designed $\ell = \pm 1$ OAM state. The histogram of mode purity is plotted in Fig. \ref{fig:4}(h), showing 49.9\% and 38.5\% purity of $\ell = \pm 1$ state. Fig. \ref{fig:4}(f) presents the reconstructed hologram, forming a "+" pattern. To systematically assess the robustness of our design, we fabricated devices with varying dimensions and grating orders. Further, to demonstrate the predictability of the cavity modes, we show in Supplementary Note 7 that their wavelengths exhibit a well-defined scaling with the hole radius.

Last, we compare our source with the state-of-the-art single-photon sources in Table \ref{tab:table1}. To the best of our knowledge, our source represents the first demonstration of hologram down to the single quanta level, which is particularly appealing to the pursuit of a quantum network with larger information capacity.

\vspace{5mm}

\section{Discussion}
In conclusion, we experimentally demonstrate integrated GP meta-cavities monolithically coupled to solid-state single-photon emitters in a CMOS-compatible GaAs platform. This approach uniquely enables simultaneous, deterministic control of single-photon properties including wavelength, spatial profile, and polarization states within an ultra-thin (\SI{200}{nm}) active layer. Crucially, these advanced manipulations are achieved without sacrificing Purcell enhancement, as evidenced by our measured Purcell factor of 9.7.

Our work establishes a paradigm for functional multiplexing: the same meta-atom geometry intrinsically encodes both the high-Q photonic cavities (traditionally requiring complex photonic crystals) and the wavefront-shaping capabilities of meta-surfaces. This dual functionality eliminates the need for heterogeneous integration of distinct photonic elements. The methodology is scalable and readily to be extended to diverse quantum light shaping regimes at the single photon level. By unifying multi-functional photon control with efficient emission in a nanoscale form factor, this strategy paves the way for large-scale, monolithic integration of high-performance quantum light sources for advanced photonic quantum technologies.

\begin{table}[h]
\caption{\label{tab:table1} Performance comparison of state-of-the-art single-photon sources with meta-surfaces.}
\begin{ruledtabular}
\begin{tabular}{cccccc}
\shortstack{Quantum \\ emitter} &
\shortstack{Extraction \\ efficiency} &
\raisebox{1ex}{\textrm{$g^{(2)}(0)$}} &
\raisebox{1ex}{\textrm{Visibility}} &
\shortstack{Purcell \\ factor}&
\shortstack{Wavefront \\ control}\\
\colrule
NV center\cite{huang2019} & -- & 0.175 & -- & No & Yes\\
GeV center\cite{komisar2023} & -- & 0.16 & -- & No & Yes\\
hBN\cite{li2023} & $\sim$25\% & 0.05 & -- & No & Yes\\
QD\cite{bao2020} & 25\% & 0.2 & -- & No & Yes\\
QD(This work) & 31.1\% & 0.017 & 0.865 & 9.7 & Yes\\
\end{tabular}
\end{ruledtabular}
\flushleft
\footnotesize\textit{Note:} NV, Nitrogen-Vacancy. GeV, Germanium-Vacancy. \\hBN, hexagonal Boron Nitride.
\end{table}

\begin{acknowledgments}
Y.C. thank Wei Liu, Weimin Ye for helpful discussions, Feng Liu for providing a test sample and Biao Yang for sparking the collaboration. This work is supported by National Key Research and Development Program of China (2022YFA1205101), National Science Foundation of China (12274296, 12192252, 12374476, 12474375), Shanghai International Cooperation Program for Science and Technology (22520714300), Shanghai Jiao Tong University 2030 Initiative. B.W. is sponsored by Yangyang Development Fund. E.H. acknowledges financial support from the Israel Science Foundation (grant number 1170/20). Y.C. is sponsored by the Innovation Research Foundation of NUDT.
\end{acknowledgments}

\bibliography{references}

\end{document}